\documentstyle[12pt]{article}
\begin{document}

\begin{titlepage}
\begin{flushright}
SNUTP--94--72   \\ August 1994
\end{flushright}
\vspace{0.4in}
\begin{center}
{\Large \bf  Search for $\eta_{c}^{'}$ and $h_{c} (^{1}P_{1})$ states \\
in the $e^+ e^-$ annihilations \\ }
\vspace{0.8in}
{\bf Pyungwon Ko \footnote{(pko@phyb.snu.ac.kr)}} \\
\vspace{0.4in}
{\sl Department of Physics  \\ Hong-Ik University \\
Seoul 121-791, Korea \\
}
\vspace{0.4in}
{\bf   Abstract  \\ }
\end{center}

\noindent
Productions and decays of spin-singlet $S,P-$wave charmonium states,
$\eta_{c}^{'}$ and  $h_{c} (^{1}P_{1})$, in the $e^+ e^-$ annihilations are
considered in the QCD multipole expansion with neglecting  nonlocality in
time  coming from the color-octet intermediate states.
Our approximation is opposite  to the Kuang-Yan's model.
The results are  $B (\psi^{'} \rightarrow h_{c} + \pi^{0}) \approx
0.3 \%$, $B ( \psi^{'} \rightarrow \eta^{'} + \gamma ) \approx 0.34 \%$,
$\Gamma (\eta_{c}^{'} \rightarrow J/\psi + \gamma) \approx 0.26$ keV
and $\Gamma (h_{c} \rightarrow J/\psi + \pi^{0}) \approx 2.5$ keV.

\vspace{0.4in}
\end{titlepage}

Among the heavy quarkonium states, the spin-singlet $P-$wave state
($h_{c,b} (^{1}P_{1})$) is extremely difficult to be produced in the
$e^+ e^-$ annihilations, because it has $J^{PC} = 1^{+-}$.
Such a state in the charmonium family was first
discovered in $p \bar{p}$ annihilation by E760 Collaboration at Fermi Lab
in 1992 (almost twenty years after the discovery of charm quark)  with the
subsequent decay $h_{c} \rightarrow J/\psi \pi^0$ \cite{e760} :
\begin{eqnarray}
\Gamma ( h_{c} ) & \sim & 800~~{\rm keV},
\nonumber \\
\Gamma ( h_{c} \rightarrow J/\psi + \pi^{0} ) & \sim & O(1)~~{\rm keV},
\label{eq:data}  \\
\Gamma ( h_{c} \rightarrow J/\psi + \pi \pi ) & < & 0.18 ~~
\Gamma ( h_{c} \rightarrow J/\psi + \pi^{0} ).
\nonumber
\end{eqnarray}
The data supports Voloshin's approach \cite{v1} instead of Kuang-Yan's
model \cite{yan} as  discussed in Refs.~ \cite{e760} and \cite{ko}.
Another spin-singlet S--wave state $\eta_{c}^{'} $ can be reached in the
$e^+ e^-$ annhilation via a spin-flip radiative decay $\psi^{'} \rightarrow
\eta_{c}^{'} + \gamma$.  However, this state has not been confirmed  yet.
Various potential models predict $m(\eta_{c}^{'})$ to be around
$\approx 3.6$ GeV.

Since BES (Beijing Spectrometer) is planning to  search for $h_{c}$ and
$\eta_{c}^{'}$ in the  decay channels   $\psi^{'} \rightarrow \eta_{c}^{'}
\gamma, h_{c} \pi^{0}$ and $h_{c} \pi \pi$, it is worth while to estimate
the  branching ratios of these decays in the frameworks of QED and QCD
multipole expasions, and compare the results with predictions of the
Kuang-Yan's model as well as with the data from BES  and other experiments.
This is precisely the purpose of this report.

First,  we discuss $E1$ and $M1$ raditaive transitions of charmonium states,
$\psi^{'}
\rightarrow  \chi_{cJ} + \gamma, \chi_{cJ} \rightarrow J/\psi + \gamma$ and
$\psi^{'} \rightarrow \eta_{c} + \gamma$,
 in order to extract various matrix elements such as $ \langle
1P | r | nS \rangle $ with $n=1,2$ and $\langle 1S | r^{2} | 2S \rangle$
from  the experimental data.  These matrix elements enter various hadronic
transitions between charmonium states that will be considered later in
this work.  We also estimate the branching ratio for $\psi^{'}
\rightarrow \eta_{c}^{'} + \gamma$.
After discussing radiative transitions of charmonium states,
we briefly recapituate how to describe hadronic transitions
between heavy quarkonia in  QCD multipole expansion and make a simple
approximation which ignores nonlocality in  time due to the color--octet
intermediate state between subsequent emissions  of gluons.
One can estimate the size of the color--octet Green's function
($G_{8}(m_{\psi^{'}})$) from the
measured decay rate for $\psi^{'} \rightarrow J/\psi + \pi\pi~ ({\rm or}~
\eta)$. For this purpose, one has to use the matrix element
$\langle 1S | r^{2} | 2S \rangle$ determined
from $\psi^{'} \rightarrow \eta_{c} + \gamma$.
One can use this $G_{8}$ to compute other decays such as $h_{c} \rightarrow
J/\psi + \pi^{0} ({\rm or}~ \pi\pi)$ and
$\psi^{'} \rightarrow h_{c} + \pi^{0}~ ({\rm or}~ \pi\pi)$.
Our results are summarized at the end.

It should be emphasized that one cannot make precise predictions for
decay rates of hadronic transitions between heavy quarkonia because of
our ignorance of color confinement in QCD.  Any attempts to estimate
such quantities  are admittedly  model dependent.   Kuang-Yan's model
takes into account  the nonlocality of $G_8$ in time, using the string
excitations near the flavor threshold.  Our approximation will be opposite
to Kuang-Yan's model in the sense that we ignore nonlocality  of $G_8$
in time, and regard $G_8$ as a constant.  Predictions of both apporaches
are a kind of rough estimates, and the actualty may lie somewhere between
two approaches.  In this sense, our predictions can be considered as
complementary to Kuang-Yan's model.


In order to describe hadronic transitions between heavy quarkonia, we have to
know such matrix elements as $\langle nS | r | 1P \rangle$ and $\langle 1S |
r^{2} | 2S \rangle$.   In previous estimations, these quantities were
replaced by the radius of the initial or the final quarkonium.
However, such estimates can be improved, since these matrix elements are
also relevant to electric dipole and spin-flip radiative transitions bewteen
heavy quarkonia.
In this work, we work in the nonrelativistic  quantum mechanics, neglecting
the relativistic corrections, $LS$ coupling and hyperfine splittings.
Therefore, the results may be affected by a factor of $\sim 2$ due to such
neglected effects.

Let us first consider  electric dipole transitions, whose  decay rates are
described by
\begin{equation}
\Gamma (E1) = {4 \over 27} \alpha \ e_{q}^{2} ~\omega^{3} ~S_{if}~
( 2 J_{f} + 1 )~| \langle n S | r | m P \rangle |^{2}.
\label{e1}
\end{equation}
Here, $J_f$ is the spin of the final quarkonium, and $\omega$ is the
energy of the emitted photon.  $S_{if} = 3$ for $h_{c}(^{1}P_{1}) \rightarrow
\eta_{c} + \gamma$ and $\eta^{'} \rightarrow h_{c} (^{1}P_{1}) + \gamma$,
whereas $S_{if} = 1$  for other transitions.

Comparing (\ref{e1}) and the measured decay rates of $\chi_{cJ} (1P)
\rightarrow J/\psi + \gamma$ and
$\psi^{'} \rightarrow \chi_{cJ} (1P) + \gamma$,    we get
\begin{eqnarray}
| \langle 1S | r | 1P \rangle_{\psi} | \approx 1.58 ~{\rm GeV}^{-1}
& = & 0.3 ~{\rm fm},
\\
| \langle 2S | r | 1P \rangle_{\psi} | \approx 1.97 ~{\rm GeV}^{-1}
& = &  0.4 ~{\rm fm},
\end{eqnarray}
where we have taken the weighted average over $J=0,1,2$ states.
Using $m(h_{c}) = 3526$ MeV and taking $m(\eta_{c}^{'}) = 3600$ MeV as
a tentative value, we get
\begin{eqnarray}
\Gamma (h_{c} \rightarrow \eta_{c} + \gamma) & \approx & 460~{\rm keV},
\label{hctogth}
\\
\Gamma (\eta_{c}^{'} \rightarrow h_{c} + \gamma) & \approx & 6.5~{\rm keV},
\label{psi2hcgth}
\end{eqnarray}
Since $| \langle nS | r | 1P \rangle_{\psi} |$'s with $n=1,2$ are
determined from the experimental data on $\Gamma (\chi_{cJ} \rightarrow
J/\psi + \gamma)$ and $\Gamma (\psi^{'} \rightarrow \chi_{cJ} (1P) +
\gamma)$,  our predictions (\ref{hctogth}) and (6) are
independent of  specific potential models.

Next, we consider a magnetic dipole radiative transition, $n ^{3}S_{1}
\rightarrow m ^{1}S_{0} + \gamma$.
The decay rate for such a spin-flip radiative decay is
given by
\begin{eqnarray}
\Gamma ( \psi^{'} \rightarrow \eta_{c} + \gamma ) & = & {4\over 3}~
\alpha Q_{c}^{2}~{\omega^{3} \over m_{c}^2}~| \langle 1S | j_{0} (\omega r/2)
| 2S \rangle |^{2}
\\
& \approx & {4 \over 3}~\alpha Q_{c}^{2}~{\omega^{3} \over m_{c}^2}~
| \langle 1S | {\omega^{2} r^{2} \over 24} | 2S \rangle |^{2}.
\end{eqnarray}
In (7), $j_{0} (x)$ is the $0-$th order spherical Bessel function,
$ j_{0} (x) = \sin x / x$.  We have used the long-wavelength approximation
($\omega r << 1$) in (8).
 From the measured decay rate $\Gamma ( \psi^{'} \rightarrow \eta_{c} +
\gamma) = 0.78 \pm 0.19$ keV, one can extract
\begin{equation}
| \langle 1S | r^{2} | 2S \rangle | \approx 2.55~~{\rm GeV}^{-2}
\label{1sr22s}
\end{equation}
which satisfies the bound (20) derived from (19).
The main point of this work is the following : one can get the matrix
element $| \langle 1S | r^{2} | 2S \rangle |$ from $\psi^{'} \rightarrow
\eta_{c} + \gamma$.
This matrix element  will be used later in order to extract $G_{8}$ from
$\psi^{'} \rightarrow J/\psi + \eta$, as described in the following.

Now, let's consider  the possibilty of finding
$\eta_{c}^{'}$ state via a radiative transition $\psi^{'} \rightarrow
\eta_{c}^{'} + \gamma$.  Assuming $m(\eta_{c}^{'}) = 3.60$ GeV, the energy
of the emitted photon is $\omega = 84$ MeV, and the matrix element in (7)
can be
approximated as $1$.   $\Gamma (\psi^{'} \rightarrow \eta_{c}^{'} + \gamma)
\approx 0.94$ keV, or $3.4 \times 10^{-3}$ in the branching ratio.
Thus, one would get $\sim 340~  \eta_{c}^{'}$'s  among $10^5~\psi^{'}$'s.
The subsequent decay of $\eta_{c}^{'}$ is mainly hadronic through
annihilation into two gluons : $\Gamma ( \eta_{c}^{'} \rightarrow 2 g )
\approx 1-4$ MeV.
The decay rate of the spin--flip radiative transition  $\eta_{c}^{'}
\rightarrow J/\psi + \gamma$ is about $0.26$ keV, using the matrix element
$|\langle 1S| r^{2} | 2S \rangle |$ which was determined above,
(\ref{1sr22s}).



Let us now discuss our main subject, the hadronic transitions between heavy
quarkonium states in QCD multipole expansion \cite{v1} \cite{yan} \cite{vz}.
The basic formulae can be found in previous works, and we write down the
relevant equations only here.
The decay rates for spin--flip transitions such as
$1^{1}P_{1} \rightarrow 1^{3}S_{1} + \pi^{0}$ and
$n^{3}S_{1} \rightarrow 1^{1}P_{1} + \pi^{0}$ are given by \cite{v1}
\begin{equation}
\Gamma (1^{1}P_{1} \rightarrow 1^{3}S_{1} + \pi^{0} )
=\Gamma (n^{3}S_{1} \rightarrow 1^{1}P_{1} + \pi^{0} )
= {1\over 2 \pi}~(A_{0}\ I_{SP})^{2} ~|\vec{p}_{\pi}|.
\label{ghtopi}
\end{equation}
Here, $A_{0} = 1.7 \times 10^{-3}~~{\rm GeV}^3$ is one-third of the matrix
element
$\langle \pi^{0} | \pi \alpha_{s} {\bf E}^{a}(0) \cdot {\bf B}^{a}(0) |
0 \rangle$, and  the matrix element $I_{SP}$ is defined as
\begin{equation}
I_{SP} = - { 2 \sqrt{3} \over 9 m_{Q}}~\langle \ nS \ | \ G_{8} (E) \ r +
r\ G_{8} (E) \ | \ 1P \ \rangle,
\label{isp}
\end{equation}
where $G_{8} (E)$ is the Green's functions
for the color octet $Q \bar{Q}$ intermediate states :
\begin{equation}
G_{8} (E) = \sum_{k} ~ { | \ k \ \rangle \ \langle \ k \ | \over E_{k} - E }.
\label{greenfunction}
\end{equation}
To get the absolute decay rate for hadronic transitions between quarkonia,
it is imperative to know more about ``$I_{SP}$'' defined in (\ref{isp}).
However,  it lies  beyond our ability to calculate $I_{SP}$ from the first
principle in QCD, because of our ignorance of the confinement in QCD.
Therefore, we have to make some reasonable approximations.
We take an opposite limit to the Kuang-Yan's approach, namely that we ignore
nonlocality in time coming from the color-octet Green's function, $G_{8}(E)$.
We simply  assume that $G_{8}(E)$ is  a constant with dimension of
inverse mass as in Ref.~\cite{ko}, since this assumption does not lead to any
contradictions to exsiting data :
\begin{equation}
G_{8} (E) = G_{8} = {\rm  constant}.
\label{greenapprox}
\end{equation}
Under this  assumption,  we have
\begin{equation}
I_{SP} = - {4 \sqrt{3} \over 9 m_{Q}}~G_{8}~\langle nS | r | 1P \rangle,
\end{equation}
and one can use the informations on the matrix elements on $r$, Eqs.~(3) and
(4).

Taking $m_{c} = 1.65$ GeV and using (3) and (4), we get
\begin{eqnarray}
\Gamma (h_{c}(1P) \rightarrow J/\psi + \pi^{0}) & = &
0.095 ~\left( {1.65\over m_{c} ({\rm GeV})} \right)^{2}~
| G_8 ({\rm GeV}^{-1}) |^{2} ~ {\rm keV},
\label{ghctopi}
\\
\Gamma (\psi^{'} \rightarrow h_{c} (1P) + \pi^{0} ) & = &
0.032 ~\left( {1.65\over m_{c} ({\rm GeV})} \right)^{2}~
| G_8 ({\rm GeV}^{-1}) |^{2} ~ {\rm keV}.
\label{gpsi2tohcpi}
\end{eqnarray}
Note that (\ref{ghctopi})  is smaller than the measured value (1) by an order
of magnitude if we assume $G_{8} \sim 1~{\rm GeV}^{-1}$, as in Ref.~
\cite{volo2}.

In order to get informations on $G_{8}(E)$, let us consider another hadronic
transition,  $\psi^{'} \rightarrow J/\psi + \eta$,
which occur through $E1-E1$ and $E1-M2$ transitions in
QCD multipole expansion \cite{yan} \cite{vz}.
Its amplitude is proportional to the following matrix element
\begin{equation}
I_{SS} = {2\over 9}~\langle \ mS \ | \ r_{i} \ G(E) \ r_{i} \ | \ nS \
\rangle,
\label{iss}
\end{equation}
which is simplified to the following under the  assumption (13) on $G_{8}(E)$,
\begin{equation}
I_{SS} = {2\over 9}~G_{8}~\langle mS | r^{2} | nS \rangle.
\end{equation}
Therefore, once $\langle 2S | r^{2} | 1S \rangle$ is known, the absolute
decay rates for  these decays can be readily obtained.
However,  potential model calculation of this matrix element is not
available on the contrary to  dipole matrix elements $\langle f \ | \ r
\ | \ i \rangle$ and the mean square radius of  a quarkonium, $\langle i \
| \ r^{2} \ | \ i \rangle$.
In Ref.~\cite{ko}, I have  used a quantum mechanical sum rule
to derive
\begin{equation}
| \langle 1S \ | \ r^{2} \ | \ nS  \rangle |^{2} < {4 \over m_{Q}}~
{| \langle 1S \ | \ r^{2} \ | \ 1S  \rangle | \over ( E_{nS} - E_{1S} )}.
\label{sumruleone}
\end{equation}
 From $| \langle 1S | r^{2} | 1S \rangle_{\psi}| = 4~~{\rm GeV}^{-2}$,
the bound  (19) was found to be
\begin{equation}
| \langle 1S | r^{2} | 2S \rangle_{\psi} |^{2} <  16.5 ~~{\rm GeV}^{-4}.
\label{sumruletwo}
\end{equation}
By considering $\psi^{'} \rightarrow J/\psi + \eta $  and using  (20),
one obtains a lower bound on $G_{8}$ \cite{ko} :
\begin{equation}
| G_{8} |^{2} > 10.4~{\rm GeV}^{-2}.
\end{equation}
Assuming $G_{8}(m_{\psi^{'}}) \approx G_{8} (m_{h_c})$, we get  the
following lower bound on $h_{c} \rightarrow J/\psi + \pi^0$ from (15) :
\begin{equation}
\Gamma (h_{c} \rightarrow J/\psi + \pi^0  ) > 1.0 ~{\rm keV},
\label{hctopiko}
\end{equation}
This is consistent with E760 data (\ref{eq:data}), and moreover, very
close to it.  Of course, this would be uncertain by a factor of $\sim 2$,
depending on  the choice of $m_c$.  (In Ref.~\cite{ko}, the mass of $h_{c}$
was assumed to be 3.51 GeV, and thus numerical results obtained there are
slight different  from the results in the present work.)

One of the main points of this work is that one can actually {\it improve}
the lower bound (22)  obtained in Ref.~\cite{ko},
using the estimate on $\langle 1S | r^{2} | 2S \rangle$ obtained from
$\psi^{'} \rightarrow \eta_{c}^{'} + \gamma$.
 From  (\ref{1sr22s})  and $\Gamma ( \psi^{'} \rightarrow J/\psi + \eta)
= ( 7.51 \pm 1.41 )$ keV, one obtains
\begin{equation}
| G_{8} |^{2} \approx 26.4~~{\rm GeV}^{-2}.
\label{g8new}
\end{equation}
Using this $|G_{8}|^2$ in (17) and (18), we finally get
\begin{eqnarray}
\Gamma ( h_{c} \rightarrow J/\psi + \pi^{0}) & \approx & 2.5~~{\rm keV},
\\
\Gamma ( \psi^{'} \rightarrow h_{c} + \pi^{0}) & \approx & 0.84~~{\rm keV},
\end{eqnarray}
Note that our prediction for $B ( \psi^{'} \rightarrow h_{c} + \pi^{0})
= 0.3 \%$ is very close to the current upper limit,
and should be checked in the near future at BEPC.
Also, these two predictions in our approach are remarkably in accord with
predictions by the Kuang--Yan's model with $\alpha_{M} \simeq 10 \alpha_E$.
However, we have only one gauge coupling (namely, $\alpha_{M} = \alpha_E$)
in our   model, and this agreement of two approaches must be regarded
accidental.

Another hadronic decay mode, $1^{1}P_{1} \rightarrow 1^{3}S_{1} + \pi \pi $,
does not receive any contribution  from the trace of the energy--momentum
tensor in QCD, and is not enhanced over $1^{1}P_{1}
\rightarrow 1^{3}S_{1} + \pi^{0}$ \cite{v1}.
(There is no anologous decay $\psi^{'} \rightarrow h_{c} + \pi \pi$ because
the phase space is not avaiable.)
Using the result of Ref.~\cite{v1}, we predict
\begin{equation}
{\Gamma (h_{c} \rightarrow J/\psi + \pi \pi ) \over
\Gamma (h_{c} \rightarrow J/\psi + \pi^0)} \approx { \lambda^{2} \over 30}
\approx 0.16,
\label{ratio}
\end{equation}
Here, $\lambda$ measures the gluonic contributions to the energy-momentum
of a pion.
This is consistent with E760 data,
and very close to the current upper limit for $\lambda \sim 2$.
Therefore, $h_{c} \rightarrow J/\psi \pi \pi$ may be observed in the near
future.

A remark is  in order.  As mentioned before, we have made an
approximation  treating the color-octet Green's function $G_{8}(E)$ as
a constant, and determined it from the measured decay rate for
$\psi^{'} \rightarrow J/\psi + \eta$.  One can actually check the consistency
of this approximation making the following observation.
The ratio
\begin{equation}
{\Gamma (h_{c} \rightarrow J/\psi + \pi^{0}) \over
 \Gamma (h_{c} \rightarrow \eta_{c} + \gamma )} =  2.1 \times 10^{-4}~
| G_{8} ({\rm GeV}^{-1}) |^{2}
\label{ratiog8}
\end{equation}
depends on $| G_{8}^2 |$, which can be used as an independent
determination  of $| G_{8}^2 |$ from the measurements.
Another ratio between $\Gamma (\psi^{'} \rightarrow h_{c} + \pi^{0})$ and
$\Gamma (\psi^{'} \rightarrow \chi_{cJ} (1P) + \gamma)$ can be useful
as well.  Thus, our assumption on $G_{8}$ is simple, does lead to
predictions  on various hadronic transitions between heavy quarkonia,
and can be checked through the measurement of the ratio (27).



In conclusion, productions and decays of $\eta_{c}^{'}$ and  $h_{c}$ in
the $e^+ e^-$ annihilations through $\psi^{'}$ decays  are
considered in the framework of QCD multipole expansion, assuming the
Green's function for the color octet states is  a constant.
Using the matrix element $| \langle 1S | r^{2} | 2S \rangle |$ extracted
from the spin--flip radiative  transition $\psi^{'} \rightarrow \eta_{c}
+ \gamma$,  we could estimate the size of $|G_{8}|^2$ by fitting
$\psi^{'} \rightarrow J/\psi + \eta$.  This leads to  predictions of
absolute decay rates,  $B(\psi^{'} \rightarrow
\eta_{c}^{'} + \gamma) \approx 0.34 \%$, $\Gamma (\eta_{c}^{'} \rightarrow
J/\psi +
\gamma) \approx 0.26$ keV, and   (24), (25),
which improves the lower bounds on the decay rates for the same processes
obtained in Ref.~\cite{ko}.
We also have suggested that our assumption (\ref{greenapprox}) could be
checked by measuring  the ratio (27).
We hope high statistics experiments at BEPC and other places searching for
$\eta_{c}^{'}$ and $h_c$ can test predictions made in this work.
Having $\sim 10^5~\psi^{'}$ decays would be enough to get the spin-singlet
$\eta_{c}^{'}$ and $h_{c}$ states and their subsequent decay properties.

\vspace{.2in}

{\Large \bf Acknowledgements \\}

The author thanks Prof. S.F. Tuan for discussions on this subject
while I was visiting  University of Hawaii at Manoa.
He is also grateful to  CTP where this work has been carried out.
This work is supported in  part by KOSEF through CTP at  Seoul National
University, and by the Basic Science Research Institute Program, Ministry
of Education, 1994, Project No. BSRI--94--2425.


\end{document}